# Simulations of an energy dechirper based on dielectric lined waveguides

Y. Nie[1], G. Xia[2,3], T. Pacey[2,3]

[1] *CERN, CH-1211 Geneva 23, Switzerland*

[2] *School of Physics and Astronomy, University of Manchester, Manchester, United Kingdom*

[3] *The Cockcroft Institute, Sci-Tech Daresbury, Daresbury, Warrington, United Kingdom*

**Abstract:** Terahertz frequency wakefields can be excited by ultra-short relativistic electron bunches travelling through dielectric lined waveguide (DLW) structures. These wakefields can either accelerate a witness bunch with high gradient, or modulate the energy of the driving bunch. In this paper, we study a passive dechirper based on the DLW to compensate the correlated energy spread of the bunches accelerated by the laser plasma wakefield accelerator (LWFA). A rectangular waveguide structure was employed taking advantage of its continuously tunable gap during operation. The assumed 200 MeV driving bunch had a Gaussian distribution with a bunch length of 3.0 μm, a relative correlated energy spread of 1%, and a total charge of 10 pC. Both of the CST Wakefield Solver and PIC Solver were used to simulate and optimize such a dechirper. Effect of the time-dependent self-wake on the driving bunch was analyzed in terms of the energy modulation and the transverse phase space.

**Keywords:** Dielectric wakefield accelerator; Ultra-short electron bunch; Energy spread dechirper.

**1. Introduction**

Beam driven dielectric lined waveguide (DLW) structures have multiple applications in the field of accelerator physics [1-4]. Among others, energy modulation introduced by the self-wake of the driving bunch can be used to compensate its time-dependent energy spread (dechirper) or produce bunch trains indirectly. Such a passive wakefield-based dechirper may simplify the linac beamline design significantly and improve the performance of the free electron lasers (FELs). The DLW-based dechirper was suggested in [5, 6] and experimentally demonstrated in [7, 8]. Corrugated metallic structure can also work as a dechirper, as proposed in [9] and demonstrated in [10].

The DLWs have been widely studied both theoretically and experimentally making use of the AWA facility at ANL [11, 12], the ATF facility at BNL [13, 14], and the FACET facility at SLAC [15, 16]. There are many other dedicated facilities being constructed or planned to study advanced electron accelerator concepts including the DLWs, for example, CLEAR at CERN [17, 18], CLARA in Daresbury Laboratory [19], SINBAD at DESY [20, 21], SPARC_LAB at INFN [17, 22], etc. These facilities can generally provide electron beams with a broad range of bunch parameters and time structures depending on operating mode. In Table 1, beam parameters are listed for CLARA (ultra-short pulse) [19] and state-of-art LWFA (laser plasma wakefield accelerator) [23-25]. Note that these parameters are not simultaneously achievable and are subject to be upgraded. As summarized in [24], typical electron beams from the LWFA have energies of a few hundred MeV (up to 4 GeV), charges of a few tens of pC, a few fs durations, relative energy spreads of a few percent (1% at best) and normalized emittances below 1 mm mrad. These energy spreads are one to two orders of magnitude larger than desired for FELs and colliders [23].

As previously studied in [21], the DLW structure has the potential to serve as a dechirper to compensate the positive energy chirp of a LWFA-accelerated bunch, in which the electrons in the bunch head have lower energy than those in the tail. In this paper, we report detailed numerical simulations of an energy dechirper that is based on a rectangular (slab-symmetric) DLW structure, with the help of the CST Wakefield Solver and PIC Solver [26].

**2. Wakefield simulations**

There are two commonly used DLW structures, the cylindrical and the rectangular dielectric waveguides. For a dechirper in the present study, the rectangular one is employed taking advantage of its continuously tunable gap [7, 8] as shown in Fig. 1. The dielectric material is diamond, due to its low RF loss tangent, high thermal conductivity and high RF breakdown threshold. The relative permittivity of diamond $\varepsilon_r$ is approximately 5.7 in THz frequency range [27].

Table 1 Beam parameters for CLARA, LWFA and PIC (particle-in-cell) simulation in the present study.

| Name | CLARA | LWFA | PIC simulation |
|---|---|---|---|
| Status | Ultra-short pulse | State of art | Present study |
| Energy (MeV) | 250 | ≤4×10$^3$ (at 6 pC) | 200 |
| Energy spread (%) | 1 | ≥1 (at 10 pC, 200 MeV) | 1 |
| Bunch length (fs) | 25 | few fs | 10 (3 μm) |
| Bunch charge (pC) | 20-100 | ≤500 (at 250 MeV) | 10 |
| Normalized rms emittance (μm) | ≤1 | <1 | 0.5 |

A three-dimensional (3D) model evaluating wakefields in slab-symmetric DLWs was presented in [28], which extends the formalism reported in [29] by including all types of modes and provides more practical closed-form formulas compared to the work presented in [30, 31]. One limitation of this model is that it assumed the charge distribution was symmetric with respect to the vertical axis. The model has been benchmarked against 3D finite-difference time-domain simulations performed with VORPAL and implemented in the popular PIC beam dynamics tracking program Impact-T. In general, the modes are classified as longitudinal section magnetic (LSM) and longitudinal section electric (LSE) modes following the definition introduced in [32]. In two-dimensional (2D) limit for large height aspect ratios (wide in $x$ direction), the LSM modes dominate, allowing one to express the modes as TM$_{0n}$-like [33, 34]. In addition to the impedance matching method, an alternative analytic approach for the decomposition of Maxwell's equations using transverse operator eigenfunctions was developed to provide a direct solution of the wakefields in an inhomogeneous rectangular DLW [35]. The analytic solutions were implemented in the RECTANGULAR code and validated by a comparison with the traditional program CST. For the present study, we focus on numerical simulations of the wakefields taking advantage of mature computer codes. Therefore, CST was applied, which enables us to compare the results from its Wakefield and PIC solvers.

First, the Wakefield Solver of CST was used to calculate the wakefields excited by the driving bunch when it passed through the DLW structure. This solver assumed that the driving bunch had a Gaussian line charge distribution. Referring to typical beams from the LWFA, the following bunch parameters were used: average energy of $E$ = 100 MeV, total bunch charge $Q$ = 1 pC, bunch length $\sigma_z$ = 3 μm. Note that beam energy has less effect on the wakefield in the relativistic regime, and the wake strength can be scaled linearly from 1 pC to 10 pC. The wake strength mainly depends on the inner gap $2a$ between the two dielectric layers for a given driving bunch. The half gap $a$ has to be small enough compared with the bunch length to excite significant wakefields, which was chosen to be 15 μm. The dielectric layer was 5 μm in thickness. The longitudinal wake impedances and on axis wakefields are shown in Figs. 2 and 3, respectively. The frequency of the first wakefield mode is higher than 5 THz. The maximum longitudinal decelerating wakefield within the driving bunch is about 33 MV/m. It can be seen that the magnitude of the decelerating wakefield within the driving bunch increases linearly along the bunch from head to tail. Such a time-dependent wake is suitable for compensating the positive energy chirp, since the trailing electrons that have higher energy will lose more energy than the head ones. For the sake of comparison, the results of the cylindrical structure with an inner radius of 15 μm are plotted as well. Due to different geometrical boundaries, the wake frequency, impedance and strength are slightly higher than that of the rectangular structure.

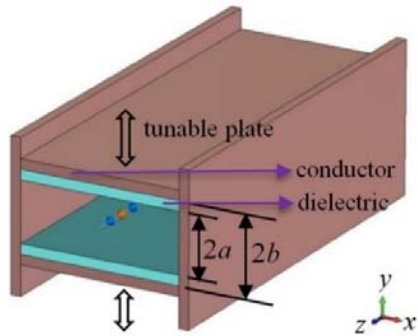

**Fig. 1.** The rectangular dielectric waveguide.

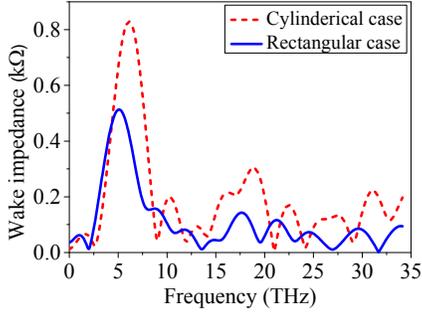

**Fig. 2.** The impedances of the wakefield. The relativistic driving bunch has a length of 3 μm and a charge of 1 pC.

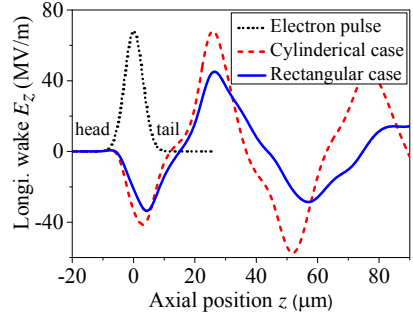

**Fig. 3.** The magnitudes of the wakefield. The relativistic driving bunch has a length of 3 μm and a charge of 1 pC.

## 3. PIC simulations

In order to understand the beam dynamics, simulations with the PIC Solver of CST have been performed as well. We assumed the output bunch parameters after the LWFA as follows: the average energy was $E$ = 200 MeV, the positive correlated rms energy spread 2 MeV, the slice energy spread ±0.3 MeV, the Gaussian bunch size $\sigma_{x,y,z}$ = 3 μm, and the bunch charge $Q$ 10 pC. Referring to [24] and Table 1, we considered these parameters would be fairly achievable for both LWFAs and traditional RF accelerators. Note that the numerical approach presented here should be easily applied to other beams. The 6D phase space distribution was generated using the code ASTRA [36] and transformed into the format required in CST by a subroutine.

Figure 4 shows the energy distributions before and after travelling through a 7.5 mm long dechirper. It can be seen that the correlated energy spread has been compensated dramatically after the dechirper while the uncorrelated energy spread (slice energy spread) has increased, since the off-axis electrons experience a varying longitudinal electric field. There is an optimal dechirper length around 7.5 mm to make the overall energy spread minimum. The energy spectrums before and after travelling 7.5 mm in the dechirper are shown in Fig. 5. The peak magnitude in the optimized energy spectrum (peak *b*) is 2.2 times the initial one (peak *a*). The energy is reduced by up to 2.5 MeV in 7.5 mm, which implies a wake strength of $E_{z,\text{dec}}$ = 333 MV/m and a dechirper strength of $S_d$ = 1.9 (MV/m)/(μm pC), where $S_d$ is the normalized wakefield strength as defined in [8], i.e., the wake strength divided by the bunch charge and the integral bunch length (about 18 μm). Normalized to the bunch charge, the wakefield strength from the PIC solver is in accordance with that from the Wakefield Solver, i.e., 33 (MV/m)/pC.

The bunch was also influenced by the transverse quadrupole wake which resulted in projected emittance growth. This quadrupole wake caused the bunch tail to be focused in the horizontal (*x*) direction while defocused in the vertical (*y*) direction. This can be seen in Fig. 6. Note that the initial normalized transverse emittance is 0.5 μm. To show the effect of the wakefield on the driving bunch, Fig. 7 plots the transverse phase spaces before and after the bunch travelling a distance of 7.5 mm both with and without the dechirper.

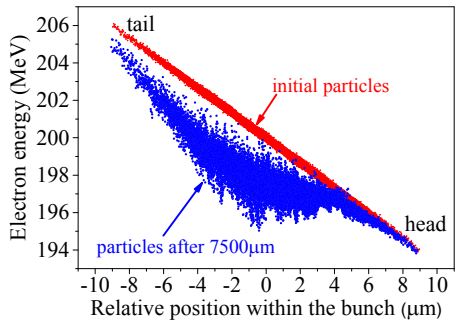

**Fig. 4.** Energy distributions before and after travelling 7.5 mm in the dechirper. The relativistic driving bunch has an rms bunch length of 3 μm and a charge of 10 pC.

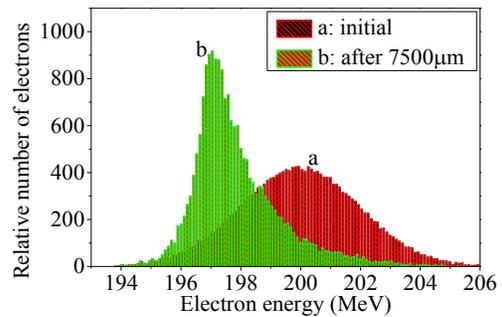

**Fig. 5.** Energy spectrums before and after travelling 7.5 mm in the dechirper. The relativistic driving bunch has a length of 3 μm and a charge of 10 pC.

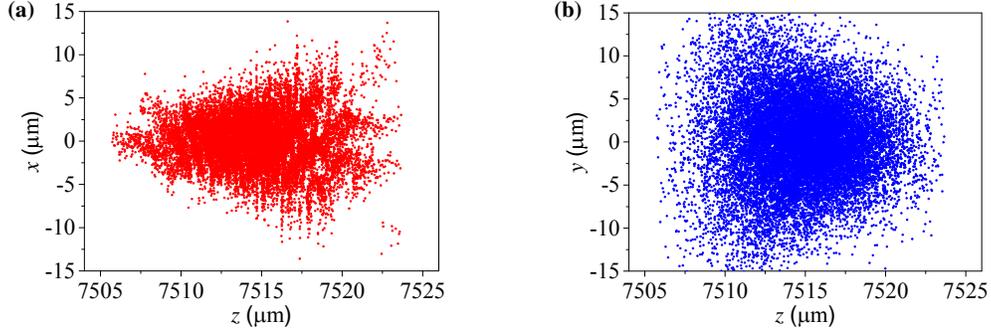

**Fig. 6.** Particle distributions in the *x-z* plane (a) and *y-z* plane (b) after travelling through the 7.5 mm dechirper. The bunch center has travelled from $z = 15$ μm to $z = 7515$ μm. Note that the initial bunch has a Gaussian distribution with a beam size of $\sigma_{x,y,z} = 3$ μm and a charge of 10 pC.

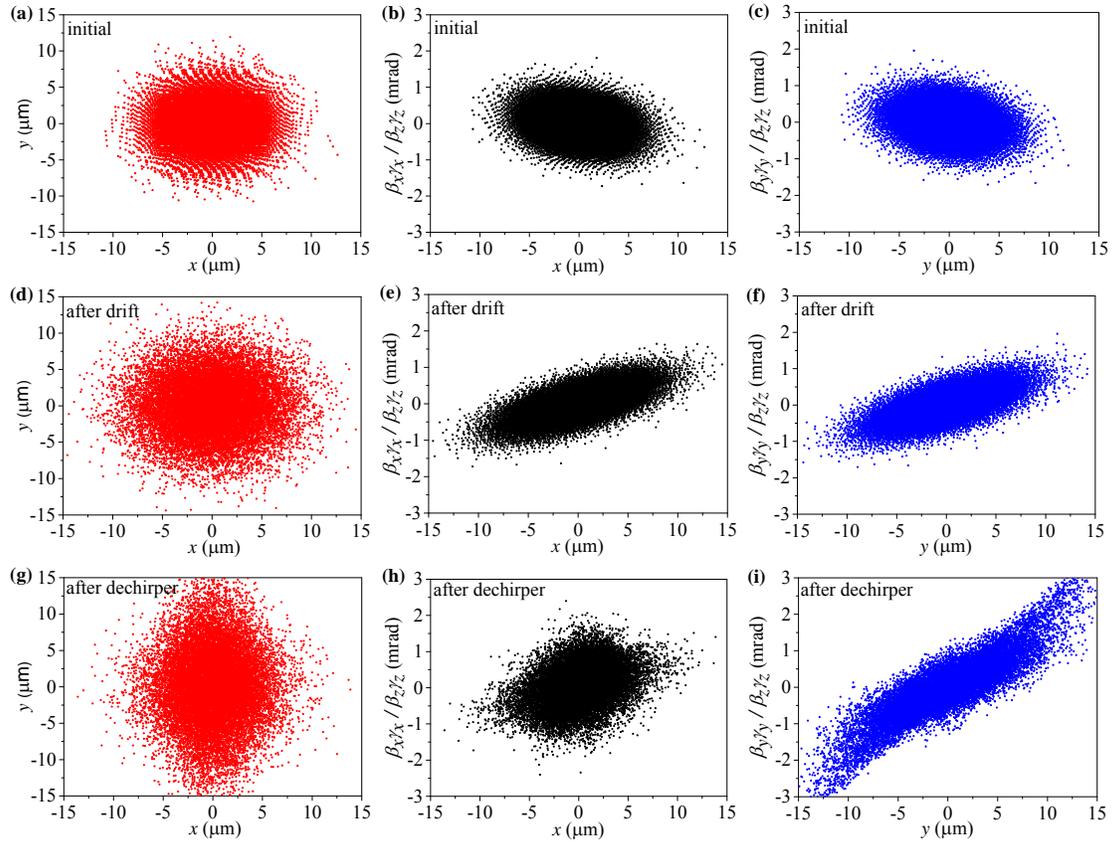

**Fig. 7.** The transverse phase spaces in front of the dechirper (a-c), after a 7.5 mm drift (d-f), and after the 7.5 mm dechirper (g-i). Note that $\beta_x\gamma_x/\beta_z\gamma_z$ denotes the ratio of the transverse momentum to the longitudinal momentum, i.e. $x' = p_x/p_z$. The same in *y*-coordinate.

## 4. Summary and discussions

Simulations have shown that the passive DLW-based dechirper has great potential to compensate the correlated energy spread of the LWFA-accelerated bunch. The advantages include its simplicity, tunability and intrinsic beam synchronization. One way to reduce the projected emittance growth is the application of a pair of rectangular DLWs with orthogonal orientations to cancel the effect of the quadrupole wake [37-40]. Shot-to-shot fluctuations of LWFA lead to a wide variety of bunches. Error tolerances need to be investigated for more realistic bunches. Essentially, if other parameters remain unchanged, energy chirp of a bunch with lower charge or bigger length than nominal will be less compensated due to weaker wakefield produced. In the relativistic regime, energy instability has less effect on the wake strength, so the level of compensation will depend on the change of absolute energy spread. Dipole wake can be excited due to pointing instability, resulting in a kick of the bunch tail and emittance growth. The transverse wakefields decrease faster than the longitudinal wakefield as the gap increases [37-39]. Therefore, we are able to adopt a larger gap to relax tolerances on beam position jitter and device alignment. Note that a longer dechirper will allow a larger gap to achieve the same effective compensation of energy chirp. On the other hand, smaller betatron function of the beam in the dechirper will also loose these tolerances and reduce the slice energy spread growth as well.

It should be noted that there are alternative ways to reduce the correlated energy spread in LWFAs, such as optical injection, ionization injection, Trojan horse injection and plasma density gradient injection. As reported in [41], the correlated energy spread was reduced by an order of magnitude making use of the beam loading effect of an escort bunch. Another promising way to provide high quality electron beams is the external injection as presented in [42].

**Acknowledgments**

The authors acknowledge Dr. C. Jing and Dr. S. Antipov from the Euclid Techlabs, USA, for the useful discussions.

**References**

[1] W. Gai, P. Schoessow, B. Cole, et al., Experimental demonstration of wake-field effects in dielectric structures, Physical Review Letters **61** (1988) 2756.
[2] Y. Nie, Wakefields in THz cylindrical dielectric lined waveguides driven by femtosecond electron bunches, Radiation Physics and Chemistry **106** (2015) 140.
[3] A. Kanareykin, Advanced acceleration and THz generation by dielectric based structures: ANL/BNL/Euclid collaboration, in: Proceedings of the 2nd European Advanced Accelerator Concepts Workshop (EAAC2015), La Biodola, Isola d'Elba, Italy, 2015.
[4] A.M. Cook, R. Tikhoplav, S.Y. Tochitsky, et al., Observation of narrow-band terahertz coherent Cherenkov radiation from a cylindrical dielectric-lined waveguide, Physical Review Letters **103** (2009) 095003.
[5] M. Rosing, J. Simpson, Passive momentum spread reduction, the wakefield silencer, ANL Report WF–144, April 1990.
[6] P. Craievich, Passive longitudinal phase space linearizer, Physical Review Special Topics – Accelerators and Beams **13** (2010) 034401.
[7] S. Antipov, C. Jing, M. Fedurin, et al., Experimental observation of energy modulation in electron beams passing through terahertz dielectric wakefield structures, Physical Review Letters **108** (2012) 144801.
[8] S. Antipov, S. Baturin, C. Jing, et al., Experimental demonstration of energy-chirp compensation by a tunable dielectric-based structure, Physical Review Letters **112** (2014) 114801.
[9] K.L.F. Bane, G. Stupakov, Corrugated pipe as a beam dechirper, Nuclear Instruments and Methods in Physics Research A **690** (2012) 106.
[10] P. Emma, M. Venturini, K.L.F. Bane, et al., Experimental demonstration of energy-chirp control in relativistic electron bunches using a corrugated pipe, Physical Review Letters **112** (2014) 034801.
[11] C. Jing, A. Kanareykin, J.G. Power, et al., Observation of enhanced transformer ratio in collinear wakefield acceleration, Physical Review Letters **98** (2007) 144801.
[12] J.G. Power, The upgraded Argonne Wakefield Accelerator facility (AWA), International Workshop on Breakdown Science and High Gradient Technology (HG2013), Trieste, Italy, 2013.
[13] S. Antipov, C. Jing, A. Kanareykin, et al., Experimental demonstration of wakefield effects in a THz planar diamond accelerating structure, Applied Physics Letters **100** (2012) 132910.
[14] ATF, <https://www.bnl.gov/atf/>.
[15] B. O'Shea, O. Williams, G. Andonian, et al., Observation of > GV/m decelerating fields in dielectric lined waveguides, in: Proceedings of LINAC2014, Geneva, Switzerland, 2014.
[16] FACET-II, <https://portal.slac.stanford.edu/sites/ard_public/facet/Pages/FACET-II.aspx>.
[17] R. Corsini, Status and outlook of the CLEAR facility, CLIC Workshop 2017, Geneva, Switzerland, 2017.
[18] R. Corsini, CALIFES description: an introduction to the proposal, its context and its timeline, CALIFES Workshop 2016, Geneva, Switzerland, 2016.

[19] G. Xia, Y. Nie, O. Mete, et al., Plasma wakefield acceleration at CLARA facility in Daresbury Laboratory, Nuclear Instruments and Methods in Physics Research A **829** (2016) 43.

[20] B. Marchetti, R. Assmann, C. Behrens, et al., Electron-beam manipulation techniques in the SINBAD Linac for external injection in plasma wake-field acceleration, Nuclear Instruments and Methods in Physics Research A **829** (2016) 278.

[21] Y.C. Nie, R. Assmann, U. Dorda, et al., Potential applications of the dielectric wakefield accelerators in the SINBAD facility at DESY, Nuclear Instruments and Methods in Physics Research A **829** (2016) 183.

[22] SPARC_LAB, <http://www.lnf.infn.it/acceleratori/sparc_lab>.

[23] A. Specka, WG1 report: Laser wakefield accelerator (LWFA), Advanced and Novel Accelerators for High Energy Physics Roadmap Workshop 2017, Geneva, Switzerland, 2017.

[24] M.E. Couprie, M. Labat, C. Evain, et al., An application of laser-plasma acceleration: towards a free-electron laser amplification, Plasma Physics and Controlled Fusion **58** (2016) 034020.

[25] C. Rechatin, J. Faure, A. Ben-Ismail, et al., Controlling the phase-space volume of injected electrons in a laser-plasma accelerator, Physical Review Letters **102** (2009) 164801.

[26] CST Software, <https://www.cst.com>.

[27] A. Kanareykin, Dielectric based accelerator: subpicosecond bunch train production and tunable energy chirp correction, in: Proceedings of the 1st European Advanced Accelerator Concepts Workshop (EAAC2013), La Biodola, Isola d'Elba, Italy, 2013.

[28] D. Mihalcea, P. Piot, and P. Stoltz, Three-dimensional analysis of wakefields generated by flat electron beams in planar dielectric-loaded structures, Physical Review Special Topics – Accelerators and Beams **15** (2012) 081304.

[29] A. Tremaine, J. Rosenzweig, and P. Schoessow, Electromagnetic wake fields and beam stability in slab-symmetric dielectric structures, Physical Review E **56** (1997) 7204.

[30] L. Xiao, W. Gai, and X. Sun, Field analysis of a dielectric-loaded rectangular waveguide accelerating structure, Physical Review E **65** (2001) 016505.

[31] C. Jing, W. Liu, L. Xiao, et al., Dipole-mode wakefields in dielectric-loaded rectangular waveguide accelerating structures, Physical Review E **68** (2003) 016502.

[32] R.E. Collin, *Field Theory of Guided Waves* (IEEE Press, New York, NY, 1991), 2nd ed.

[33] T.C. Marshall, C. Wang, and J.L. Hirshfield, Femtosecond planar electron beam source for micron-scale dielectric wake field accelerator, Physical Review Special Topics – Accelerators and Beams **4** (2001) 121301.

[34] G. Andonian, D. Stratakis, M. Babzien, et al., Dielectric wakefield acceleration of a relativistic electron beam in a slab-symmetric dielectric lined waveguide, Physical Review Letters **108** (2012) 244801.

[35] S.S. Baturin, I.L. Sheinman, A.M. Altmark, et al., Transverse operator method for wakefields in a rectangular dielectric loaded accelerating structure, Physical Review Special Topics – Accelerators and Beams **16** (2013) 051302.

[36] ASTRA code, ⟨http://www.desy.de/~mpyflo/Astra_for_WindowsPC⟩.

[37] Z. Zhang, K. Bane, Y. Ding, et al., Electron beam energy chirp control with a rectangular corrugated structure at the Linac Coherent Light Source, Physical Review Special Topics – Accelerators and Beams **18** (2015) 010702.

[38] F. Fu, R. Wang, P. Zhu, et al., Demonstration of nonlinear-energy-spread compensation in relativistic electron bunches with corrugated structures, Physical Review Letters **114** (2015) 114801.

[39] M.W. Guetg, K.L.F. Bane, A. Brachmann, et al., Commissioning of the RadiaBeam/SLAC dechirper, in: Proceedings of IPAC2016, Busan, Korea, 2016.

[40] T.H. Pacey, G. Xia, Y. Saveliev, et al., Phase space manipulation of sub-picosecond electron bunches using dielectric wakefield structures, in: Proceedings of IPAC2017, Copenhagen, Denmark, 2017.

[41] G.G. Manahan, A.F. Habib, P. Scherkl, et al., Single-stage plasma-based correlated energy spread compensation for ultrahigh 6D brightness electron beams, Nature Communications **8** (2017) 15705.

[42] M.K. Weikum, R.W. Assmann, U. Dorda, et al., Improved electron beam quality from external injection in laser-driven plasma acceleration at SINBAD, in: Proceedings of IPAC2017, Copenhagen, Denmark, 2017.